\documentstyle[12pt,aasms4]{article}
\def\ni{\noindent}
\def\spose#1{\hbox to 0pt{#1\hss}}
\def\lta{\mathrel{\spose{\lower 3pt\hbox{$\mathchar"218$}}
     \raise 2.0pt\hbox{$\mathchar"13C$}}}
\def\gta{\mathrel{\spose{\lower 3pt\hbox{$\mathchar"218$}}
     \raise 2.0pt\hbox{$\mathchar"13E$}}}

\begin{document}

\title{Some Observational Consequences of GRB Shock Models}
\author{Pawan Kumar}
\affil{Institute for Advanced Study, Princeton, NJ 08540}
\author{Tsvi Piran}
\affil{ Racah Institute, Hebrew University, Jerusalem 91904, Israel,
Physics Department, Columbia University, New York, NY, USA,
Physics Department, New York University, New York, NY, USA }
\authoremail{pk@ias.edu
tsvi@nikki.fiz.huji.ac.il}

\begin{abstract}
\baselineskip 15pt

Gamma-ray bursts are believed to be produced when fast moving ejecta
from some central source collides with slower moving, but
relativistic, shells that were ejected at an earlier time. In this so
called internal shock scenario we expect some fraction of the energy
of the burst to be carried by slow moving shells that were ejected at
late times. These slow shells collide with faster moving outer shells
when the outer shells have slowed down as a result of sweeping up
material from the ISM. This gives rise to a forward shock that moves
into the outer shell producing a bump in the afterglow light curve of
amplitude roughly proportional to the ratio of the energy in the inner
and the outer shells.  In addition, a reverse shock propagates in the
inner shell and produces emission at a characteristic frequency that
is typically much smaller than the peak of the emission from the outer
shell by a factor of $\sim 7 \gamma_{0c}^2 (E_2/E_1)^{1.1}$, and the
observed flux at this frequency from the reverse shock is larger
compared to the flux from the outer shell by a factor of $\sim 8
(\gamma_{0c} E_2/E_1)^{5/3}$; where $\gamma_{0c}$ is the bulk
Lorentz factor of the outer shell at the time of collision, and $E_1$
\& $E_2$ are the total energy in the outer and the inner shells respectively.
The Lorentz factor is related to the observer time as $\sim 5
(t/day)^{3/8}$.  The shell collision could produce initial temporal
variability in the early afterglow signal. The lack of significant
deviation from a power-law decline of the optical afterglow from half
a dozen bursts suggests that $E_2/E_1$ is small. Future
multi-wavelength observations should be able to either detect bumps in
the light curve corresponding to both the forward and the reverse
shocks or further constrain the late time release of energy in ejecta
with small Lorentz factor, which is expected generically in the
internal shock models for the gamma-ray bursts.

\end{abstract}

\bigskip
\hskip0.3cm{\it Subject headings:\rm~ gamma rays: bursts -- relativistic shock}

\vfill\eject
\baselineskip 15pt

\section{Introduction}

The recent multi-wavelength observations of GRB afterglow
(Costa et al. 1997, van Paradijs et al. 1997, Bond 1997, Frail et al. 1997)
support the fireball shock model for GRBs (Wijers et al. 1997, Vietri 1997, 
Katz \& Piran 1997, Waxman 1997). According to this model GRBs are produced 
as a result of internal shocks when a fast moving shell runs into a slower 
moving shell that was ejected at an earlier time. The relative kinetic energy 
of motion of the shells is converted into observed gamma-ray emission via a
relativistic shock. The observed afterglow emission in this scenario
is produced when the shell slows down as a result of interaction
with the ISM. This internal-external fireball model (Meszaros \& Rees 1997,
Sari \& Piran 1997a-b) 
requires a complicated central engine, which operates for a long 
time (as long as the duration of the burst) producing a highly 
variable flow. The observed temporal structure seen in GRBs, follows, 
to a large extend, the temporal structure produced by the 
source (Kabayashi, Piran \& Sari, 1997). 
M\'esz\'aros et al. (1998) considered the effect of an inhomogeneous fareball,
where the fireball parameters had angular dependence, on the afterglow.
At the end of the internal 
shock phase we are left with a rather ordered flow in which faster 
(merged) shells are the outermost ones and slower shells follow behind 
them (see fig. 1). If the source continues to operate for even longer 
time, producing even slower shells, they will follow behind the 
earlier ejected faster moving shells.

As the fast outer shells move outwards they begin to interact with the
ISM and decelerate. Eventually the slower inner shells will catch up
with the outer shells.  Thus one should expect late
interaction of slow shells with faster shells that has been slowed
down. Our goal in this paper is to examine such collisions and to
provide some observational signatures of such interactions which should
be seen in future afterglow observations. The current and the future
observations can also be used to place a limit on the fraction of energy
coming out in slow moving ejecta at late times.
As an example the hypernova model (Paczynski, 1997), in which perhaps
most of the energy comes out from the central engine in the from
of slow moving ejecta that lags behind the outer fast material, should
show deviation at late times from smooth power-law decline of the 
after-glow light curve.

Rees \& M\'esz\'aros (1998), Panaitescu et al. (1998), and 
Cohen \& Piran (1999) have recently considered the continuous limit of the
delayed emission problem focusing on the forward shock emission.
They discussed a source that emits a wind
whose characteristics varies as a power law with time. However, in
view of the internal shock model we expect that an impulsive
situation in which two shells collide should be more relevant to GRB
afterglows. We consider therefore, a toy model consisting of two shells
that undergo collision, and we calculate synchrontron emission from both
the forward as well as the reverse shock that propagates into outer and 
the inner shells respectively. The
interaction of the outer shell with the ISM is described well by the
adiabatic Blandford-McKee (1976; hereafter BM) self-similar solution.
The inner shell catches up with the outer shell when the outer shell
slows down and its Lorentz factor becomes approximately equal to the
Lorentz factor of the inner shell.  The shell collision produces two
new shock waves: a forward shock that moves into the outer shell and a
reverse shock that propagates into the inner shell. We analyze these
shocks and estimate the resulting synchrotron emission.

If the collision takes place not too early during the afterglow the
system is adiabatic (this is probably a valid approximation in GRB
afterglow from about half an hour after the burst;
Granot, Piran \& Sari, 1998) and the outer material has arranged
itself with a BM self-similar profile.  In this case the overall
effect of the collision between the outer shell, with energy $E_{1}$
and the inner shell, with energy $E_2$ would be a transition from one
BM solution, with a total energy $E_1$ to another one, with a total
energy $\sim E_1 +E_{2}$, i.e. the observed flux at a fixed frequency
will increase by a factor of $(1+E_2/E_1)^{1.4}$. This estimate is not
exact as there would be some enhanced energy losses form the shocks
that arises during the collision. At most wavelengths we should expect
a smooth transition from one solution to the other. The exception is
at those frequencies at which there is significant emission from the
reverse shock. As we show later the emission from the forward shock is
rather similar to the emission from the shock that arises from the
interaction of the outer shell with the ISM.  The total emission from
the reverse shock (the shock that propagates into the inner shell) is
smaller but typically comes out at a much lower frequency. At these
frequencies the impact of this additional emission is likely to be
very pronounced.

Early on, during the early afterglow we expect that the system is
radiative (it must be radiative during the GRB phase, otherwise the
energy budget would be much higher). In this case we expect that the
outer shells that has collided with the ISM would cool rapidly and the
collision between the inner and the outer shell would be between two
cold materials. The calculations presented in section
\ref{Shock_conditions} and \ref{outer_shell} are valid only within the
adiabatic regime, namely several hours after the beginning of the
afterglow. However, similar qualitative behavior also applies 
during the early afterglow while the shock is still radiative.
In particular, after the internal shocks have arranged the outflow in
a monotonically increasing function of radius, the successive
collisions of these shells would lead to a variability in
the early afterglow with the variability time-scale of the order of the
time since the explosion.

Note that we have made several simplifying assumptions in our
model. In addition to the above mentioned assumptions of local
spherical behaviour (which is a good approximation at this early stage
even if the system is beamed) and adiabaticity (which might be broken
as we discuss below) we have assumed that (i) the ratio of electron
and magnetic field energy density to the total thermal energy is the
same behind all shocks and (ii) that the ratio of specific enthalpy in
the inner and the outer shells is approximately given by the
ratio of their energy.  This latter
assumption is valid only if the ratio of the thickness of the slow
shell and its radius is approximately given by the inverse of the
square of the Lorentz factor.  We discuss these assumptions after
equation \ref{stam} in section \ref{synchemission} and in a footnote in
section \ref{Shock_conditions}. Clearly our results should be modified
if these assumptions are not satisfied.

We describe our physical model in section 2. A detailed discussion of 
the shock conditions, when a cold and a hot shells collide, 
is presented in \S2.1, and in \S2.2 we discuss the effect of the 
stratification of the outer shell, using the Blandford-McKee (BM) 
solution, on the shock propagation and emission. We discuss briefly
the situation in a radiative case in \S2.3
The synchrotron emission from the shocks and its observational
consequences are discussed in \S3. 
Implications to recent observations and predictions for future
observations are contained in the final section.

\section{The Physical Model}
\label{physical}

Our basic model consists of a {\it cold} inner shell that is colliding 
with an outer {\it hot} shell. The outer shell is slowing down as a result
of collision with the interstellar medium (ISM) so that the slower
moving inner shell eventually catches up with it.

Let us take the energy in the outer shell, as measured in rest frame of
the center of the explosion, prior to the shell collision, to be $E_1$, 
and the energy in the inner shell to be $E_2$. The Lorentz factor of
the outer shell, $\gamma_0(t)$, decreases with time. At late times
when $\gamma_0(t)$ is much less than the initial value, $\gamma_{0i}$,
the Lorentz factor is given by

\begin{eqnarray}\label{gamma0t}
\gamma_0(t)\approx \left({3E_1\over 16\pi\rho_0 t^3}\right)^{1/2} 
   = {\gamma_{0i}\over 2} \left({t_0\over t}\right)^{3/2}
\end{eqnarray}
where $\rho_0$ is the density of the interstellar medium, and
\begin{eqnarray}
t_0 = \left({3 E_1\over 8\pi\gamma^2_{0i}\rho_0}\right)^{1/3}
\end{eqnarray}
is the time when $\gamma_0(t)=\gamma_{0i}/2$.

The radius of the inner shell at time $t$ is $R_4(t) \approx t - 
t/(2\gamma_4^2)$, and the radius of the decelerating outer shell
is $R_0(t) \approx t - t/[8\gamma_0^2(t)]$; $\gamma_4$ is the Lorentz
factor of the inner shell, and all quantities are measured in
the rest frame of the center of the explosion. The inner shell runs 
into the outer shell when $R_4(t) = R_0(t)$, and at this time $\gamma_0(t) 
\equiv \gamma_{0c} = \gamma_4/2$, and thus the relative Lorentz factor
of collision is approximately 1.25 so long as we neglect the slowdown of 
the inner shell. The time when the shells collide is 
$t_c \approx t_0 (\gamma_{0i}/\gamma_4)^{2/3}$. 
For the self-similar structure of
the outer shell, described by the Blandford-McKee solution, it can be
shown that the Lorentz factor of the inner shell w.r.t. the rest frame
of fluid of the outer shell, when the collision takes place, is about 1.25.
Therefore, the relative speed of 
propagation of the two shocks, one propagating in the outer and the other 
in the inner shell, is time independent and is equal to 0.6 $c$. In this 
case the outgoing forward shock, that propagates into the already shocked 
material of the outer shell, is weak and the reverse shock propagating into 
the cold inner shell is strong but mildly relativistic. 

The space can be divided into five different regions.
(I) The ISM is the outermost region which consists of cold gas, is
taken to be at rest relative to distant observers, and is characterized
by a single parameter the density, $n_0$.
(II) The outer shell and the shocked ISM consists of relativistic
electrons and protons moving with a bulk Lorentz factor of $\gamma_0(t)$
relative to the ISM; the thermodynamical variables in this region
are denoted by a subscript 1 and the equation of state is $p_1 = e_1/3$.
Region III consists of material of the outer shell that has been
heated by the forward shock resulting from shell collision; this
region is also relativistic and the variables are denoted by a subscript 2.
(IV) Part of the inner shell that has been heated by the reverse shock;
electrons are relativistic in this region but the protons are not and
consequently the equation of state is $p_3 = 4 e_3/9$. (V) This region 
contains unshocked cold inner shell characterized by density
$n_4$. This shell moves with a Lorentz factor $\gamma_4$ w.r.t. 
the ISM and $\gamma_{14}$ relative to region 2. The different
regimes are depicted in Fig. 2.

\subsection{Shock conditions}
\label{Shock_conditions}
We analyze a pair of forward shock and reverse shock formed when 
a cold region ($n_4 m_p c^2 \gg e_4$) collides with a hot 
relativistic region ($n_1 m_p c^2 \ll e_1 = 3 p_1$). The system is
characterized by three parameters: the energy density in the hot
region, $e_1$, the particle density in the cold region, $n_4$, and the
relative Lorentz factor between the inner and the outer shells, 
$\gamma_{14}$. In fact most of the quantities are determined just 
by the dimensionless ratio of the enthalpy density $ w_4/w_1 = 
3 n_4 m_p c^2/(4 e_1) \sim E_2/E_1$, and by $\gamma_{14}$.
\footnote{In deriving the relation $ w_4/w_1 \sim E_2/E_1$ we made use of 
the fact that the thickness of the cold inner shell is $R_4/\gamma_4^2$ 
as seen in an inertial frame. The thickness of a shell, in shell rest frame, 
which is undergoing radial expansion at the sound speed $\sim c/3^{1/2}$, is
approximately $c t_s = c t/\gamma_4$; where $t_s$ and $t$ are the time
since the shell was expelled from the central source in the shell
reast frame and as seen by an inertial observer respectively. Thus the
shell thickness in an inertial frame is $\sim c t /\gamma_4^2$ provided
that the shell radius at the time of collision is much larger than its 
radius at expulsion, and at collision the sound speed
of the inner shell is not highly sub-relativistic. The former condition
is satisfied at all times during and after the $\gamma$-ray burst. The 
shell thickness is less than $c t /\gamma_4^2$ when the latter condition
is violated and in this case the peak frequency of emission from the reverse
shock is correspondingly smaller. } 
A fourth parameter ($n_1$) does not appear in the shock conditions 
but it determines the density in the shocked region III ($n_2$).

Two shocks form, the forward shock propagates into the hot outer shell and
the reverse shock propagates into the cold inner shell.
Each of the shocks satisfy three jump conditions which ensure 
the conservation of particle number, energy and momentum across the 
shock. For the forward shock propagating into region II (see fig. 2) 
we have:
\begin{eqnarray} 
n_1 \gamma_1 v_1 & =& n_2 \gamma_2 v_2 \\
w_1 \gamma_1^2 v_1 & =& w_2 \gamma_2^2 v_2 \\
w_1 \gamma_1^2 v_1^2 + p_1 &=& w_2 \gamma_2^2 v_2^2 + p_2
\label{jump}
\end{eqnarray}
where $\gamma_1$ ($\gamma_2$) and the corresponding velocity $v_1$
($v_2$) are measured with respect to the rest frame of the shock and
$n_i$, $e_i$, $p_i$ and $w_i$ are the particle density, thermal energy
density, pressure and enthalpy measured with respect to the rest frame
of the fluid in region `i+1'.

The forward shock propagates into hot material that has been already
shocked when the first shell interacted with the ISM. The mass energy 
density in the outer shell is negligible relative to the thermal
energy density i.e. $ n_1 m_p c^2 \ll e_1 $, and the equation of 
state is: $p_1 = e_1/3$. As the shocked material can be only hotter,
similar conditions should hold in region III i.e. $n_2 m_p c^2 \ll e_2 $ 
and $p_2 = e_2 /3$. We also expect
that $e_2 \ge e_1$.  With these conditions we find:
\begin{eqnarray}\label{forward}
v_2 & = & \frac{1}{3 v_1} \\
\gamma_1 & =& {\sqrt{{\frac{3(3e_1 + e_2)} {8 e_1}}}} \\
\gamma_2 & =& {\sqrt{{\frac{3(e_1 + 3 e_2)}{8 e_2}}}} = {\sqrt{{\frac
   {9(\gamma_1^2-1)}{(8\gamma_1^2 - 9)}}}} \\
\gamma_f & = & {\sqrt{{\frac{\left( 3e_1 + 
             e_2 \right) \left( e_1 + 3e_2 \right) }{{ 16
             e_1}e_2}}}} = {\frac{3 v_1^2-1}{2 v_1}}\\ 
n_2 & = & n_1\gamma_1 v_1^2 {\sqrt{{\frac{(8\gamma_1^2-9)}{\gamma_1^2-1}}}}
     = {\sqrt{{\frac{e_2 \left( e_1 + 3e_2 \right) }{{e_1 }\left( 3e_1 + 
    e_2 \right) }}}} n_1 \\
\gamma_{t2} & = & \frac{e_2}{n_2 m_p c^2} = \gamma_{t1} {\sqrt{{\frac
   {(8\gamma_1^2-9)\gamma_1^2}{9(\gamma_1^2-1)}}}},
\end{eqnarray}
where $\gamma_f$ is the relative Lorentz factor between regions II and III,
thus it is the shocked matter velocity in the rest frame of the unshocked
material, and $\gamma_{t2}$ is the ``thermal'' Lorentz factor of 
protons in region III or the typical Lorentz factor of the random 
motion of the protons and  is crucial in determining the emission 
from this region.

The reverse shock is different. In the region of the reverse shock
(marked IV in fig. 2) the electrons are relativistic but the protons
are non-relativistic or at best mildly relativistic. The equation of 
state in this region is $p_3 = \eta e_3/3$, where $\eta =1$ for 
relativistic protons and $\eta = 4/3$ for non-relativistic protons.
For the unshocked cold shell $e_4 = p_4 = 0$.
Separating regions III and IV is a surface of contact discontinuity. 
Both regions move with the same velocity and both have the same pressure: 
$p_2= p_3$. The energy densities in the two regions are related by
$e_2 = \eta e_3$. Using these relations and the jump conditions 
analogous to equations (3)--(5) we obtain:
\begin{eqnarray}\label{reverse}
\gamma_4 &=& {\sqrt{{\frac{\left( e_2\eta + n_3 - n_4 \right) 
        {{\left( e_2 + 3n_4 \right) }^3}}{\left(  e_2
           \left( -1 + 3\eta \right)  + 3n_3 - 3 n_4 \right) n_4}}}} \\
\gamma_3& =& 9{\sqrt{{\frac{\left( e_2 + n_3 \right) 
         \left( e_2 + n_3 - n_4 \right) }{\left( 13e_2 + 9n_3 \right) 
         \left( 5e_2 + 9n_3 - 9n_4 \right) }}}} \\
\gamma_r&=&\sqrt{\frac{\left( \eta e_2  + n_3 \right) 
     \left( e_2 + 3n_4 \right) }{\left[
        \left( 1 + 3\eta \right)  e_2 + 3n_3 \right] n_4}} \\
n_3    &=& {\frac{\left(1+6\eta \right) n_4  
 + {\sqrt{4 \eta e_2  n_4 \left(1 + 3\eta \right) 
 + {{\left(1+6\eta \right) }^2}{{ n_4}^2}}}}{2}} \\
 e_3 &= &\eta e_2 \\
\gamma_{t3}& =& (e_3 / n_3 m_p c^2) .
\end{eqnarray}
Here, $\gamma_r$ is the Lorentz factor of the motion of the shocked
matter in region IV relative to the cold matter in region V.  To avoid
cumbersome equations we have expressed $\gamma_4$, $\gamma_3$ and
$\gamma_r$ in terms of $n_3$. However, $n_3$ is given in a closed form
in terms of $e_2$, and $n_4$. So $e_2$ is the only remaining unknown
parameter in equations (7)--(17). Since regions III and IV move with 
the same velocity we can express the relative Lorentz factor between 
the cold inner shell, region V, and the outer shocked material, II, in terms 
of the Lorentz factors $\gamma_f$ and $\gamma_r$:
\begin{eqnarray}\label{gammma14} 
\gamma_{14} = \gamma_r \gamma_f + \sqrt{[(\gamma_r^2 -1)(\gamma_f^2 -1)]}
\end{eqnarray}
For given parameters ($e_1$, $n_4$ and $\gamma_{14}$) we can solve
equations (9), (14), (15) \& (18) to determine $e_2$ and thus all other 
variables in the shocked regions. Note that $e_2/e_1$ depends only 
on $\gamma_{14}$ and the ratio $n_4/e_1\propto w_4/w_1$. 

The ratio of thermal energy densities in regions III and II, $e_2/e_1$,
is a function of the relative Lorentz factor of collision ($\gamma_{14}$)
and $w_4/w_1$. For a fixed value of $\gamma_{14}\approx 1.25$ we find
from the numerical solution of the conservation equations that $e_2/e1 \approx
[1+ (w_4/w_1)^{0.35}]$. The ratio of the densities $n_2/n_1$ is an even
weaker function of $(w_4/w_1)$.

There is a region in the parameter space for which the above set of
equations have no solution. The no-solution case physically
corresponds to when the enthalpy density, or the energy, in the inner
shell is too small to drive a forward shock into the outer shell. This
is best seen in fig. 3 which depicts $e_2/e_1$ vs.  $w_4/w_1$ for
several values of $\gamma_{14}$. Since regions III is shocked, $e_2 >
e_1$ with $e_2 = e_1$ a limiting case in which the forward shock
disappears. Therefore, for a given value of $\gamma_{14}$ the minimal
value of $w_4/w_1$ is obtained when $e_2=e_1$.  The minimal value of
$w_4/w_1$ for a forward shock to occur increases with decreasing 
$\gamma_{14}$; for $\gamma_{14}=1.25$ the minimum
value of $w_4/w_1$ is 0.36.

The correct treatment of the problem in those cases where we don't have a
solution of the conservation equations is to relax the homogeneous shell 
approximation used thus far and use the Blandford-McKee solution for the 
structure of the outer shell. As $e_1$, the energy density in region II
decreases with distance from the interface of regions I \& II,
there will always be a point in region II at which $w_4/w_1$ is large 
enough so that forward shock forms. This is discussed in the next section.

\subsection{Structure of the outer shell and shock solution}
\label{outer_shell}
We turn now to determine the parameters for the outer shell (region II of 
fig. 2). This region contains shocked material from the ISM and the 
baryonic ejecta from the explosion.
There is a shock between this region and the cold ISM.
In a simple relativistic shock between two cold shells we have (see
e.g. Piran 1999):
\begin{equation}
n_1 \approx 4 \gamma_0 n_0,
\ \    \ \  e_1  = 4 \gamma_0^2 n_0 m_p c^2 .
\label{region_1}
\end{equation}

The simplest approximation would be to assume that region II is
homogeneous and to use equation (19), together with the
adiabaticity condition:
\begin{equation}
\label{adiabatic}
E_1 = (4 \pi /3) R^3 n_0 m_p c^3 \gamma_0^2 ,
\end{equation}
to determine the conditions in region II.  However, the inner shell
overtakes the outer one when the outer shell has shocked on the
ISM and has slowed down to a Lorentz factor that is roughly equal to that
of the inner shell. At this stage a rarefaction wave has propagated 
through the outer shell and it has settled down to a
Blandford--McKee (BM) self-similar solution where the density, enthalpy, bulk
Lorentz factor etc. decrease with distance from the outer surface of the shell.
The collision of the shells and the resulting structure and emission from
the shocked regions should be calculated using the stratified structure
of the outer shell given by the BM solution.

The BM solution is expressed in terms of the similarity variable
$\chi$, which is defined by:
\begin{equation}
\label{def_chi}
\chi \equiv 1+16 \gamma_f^2\left({r \over R}\right) \ ,
\end{equation}
where $R$ is the radius of the shock front, $r$ is the distance inward
from the shock from (measured in the ISM rest frame) and $\gamma_f$ is
the Lorentz factor of the matter just behind the shock:
\begin{equation}
\label{gamnmaf}
R^3 \gamma_f^2  = {17 \over 12} l^3 
\end{equation}
where $l^3 \equiv E/[(4 \pi/3) n_0 m_p c^2]$. $R$ and $\gamma_f$ are
related to the observed time at which radiation from the front of the
shock reaches the observer as:
\begin{equation}
t_{obs} = { 3 \over 68} { R^4 \over l^3 c} \approx {l \over 14 c \gamma_f^{8/3}}
\end{equation}

The BM self-similar profile is given by:
\begin{equation}
\label{BM}
n'= 4 \gamma_f n_0 \chi^{-5/4} \ \ \ ,\ \ \ 
\gamma'=\gamma_f \chi^{-1/2} \ \ \ ,\ \ \ 
e'=4 n_0 m_p c^2 \gamma_f^2 \chi^{-17/12} \ ,
\end{equation}
where $n'$ and $e'$ are the number density and the energy density in
the local rest frame, respectively, and $\gamma'$ is the Lorentz factor 
of the bulk motion of the matter at radius ($R-r)$. 

A complete and exact solution of the colliding shell problem requires
determining the propagating of shock wave through a BM stratified medium. 
This could be done numerically. However, an approximate analytic solution 
should be sufficient for our purpose. 

As discussed at the beginning of \S2 the relative Lorentz factor of collision 
i.e., the relative Lorentz factor of region V and the bottom of region 
II ($\gamma_{14}$), is 1.25 and this is independent of $\chi$. Thus, as 
the forward shock propagates outward through the stratified outer shell, 
the value of $\gamma_{14}$ remains nearly constant. However, 
the value of $w_4/w_1$ decreases as $\chi^{17/12}$; $w_4/w_1$ at $\chi=1$ 
is approximately time independent. 

In the last sub-section we found that for a fixed value of $\gamma_{14}$
the forward shock dies out if $w_4/w_1$ falls below some critical value;
for $\gamma_{14}=1.25$ the minimum value of $w_4/w_1$ is 0.36.
Thus, if the energy of the inner shell ($E_2$) is less than about 0.36 $E_1$,
the forward shock stalls and turns into a rarefaction wave when it reaches 
the surface $\chi_c \sim 0.5 (E_1/E_2)^{12/17}$. 

For $\chi_c\lta 1$, the forward shock traverses through the
outer shell and the energy density at a radius $\chi$, within the shell,
increases as a result of the shock 
by a factor proportional to $\chi^{17/48}$, and the particle number 
density increase is an even weaker function of $\chi$.
Since the synchrotron emission scales as ${e}^3/n$,
most of the emission from the outer shell, pre- as well as post-collision,
is generated near the top of the shell. The overall increase in the 
emission from the outer shell can, therefore, be calculated by treating
the shell to be homogeneous with values of $n_1$, $e_1$ and 
$\gamma_1$ corresponding to $\chi=1$ in the BM solution. Emission
from the reverse shock can also be calculated using the same values of
these parameters.

For $\chi_c\gg 1$ the forward shock is very weak, it stalls at $\chi=\chi_c$,
and has small effect on the emission from the outer shell. The emission 
from the reverse shock can however be still much larger than the emission 
from the outer shell at the frequency corresponding to the peak of the 
synchrotron radiation from region IV. This emission can be calculated 
by applying the results of the last section with $\gamma_{14}=1.25$, 
$w_4/w_1 = 0.36$, and other parameters in region II corresponding to 
$\chi=\chi_c$ of the BM solution.

\subsection{Radiative Collisions}

Slower shells that catch up with faster decelerating shells during 
the first few hours of the burst find the outer shell to be cold as 
the synchrotron cooling time is shorter than the dynamical time,
and thus we need to consider collision between two cold shells.

The calculation of shocks resulting from the collision of two cold
shells resembles the calculation of energy emitted from internal shocks
(Sari \& Piran, 1997a). It is also rather similar to the calculation
given in \S2.1. The main difference is that the pressure in region II is
zero, and as a result there is always a forward and a backward
shock. At early times, within a few hours of the burst, the shock is
radiative and the Lorentz factor drops off with time as $t^{-3}$. 
The relative Lorentz factor of the shells at the time of collision is about
1.5. Simple kinematic calculation suggest that, equal masses colliding with 
a relative Lorentz factor of 1.5 and radiating efficiently will radiate away
$\sim$10\% of their total energy. The rest of the energy will simply
be added to the kinetic energy of the ejecta plowing through the ISM.
We find that the Lorentz factor of the forward and the reverse shocks
are about 1.12 when $n_1\approx n_4$. Both shocks are similar
and they emit their energy at a much lower frequency than the
outermost shock that is propagating into the ISM at this stage. We
expect, therefore, that this emission will be a very significant
contribution to the emission at this lower wavelength at this stage.

\section{Synchrotron Emission}
\label{synchemission}
The peak of synchrotron emission occurs at frequency
$\nu_m\approx q_e\gamma_e^2 B/(m_e c)$,
and the frequency integrated emissivity is given by $\epsilon\approx
\sigma_T c n_e
\gamma_e^2 B^2/8\pi$; where $q_e$, $m_e$, $\gamma_e$ are electron charge, 
mass, and thermal Lorentz factor respectively, $B$ is the magnetic field in 
the fluid rest frame, $n_e$ is the electron number density, and $\sigma_T$ 
is the Thomson cross-section. The emission at low frequencies ($\nu\ll\nu_m$)
drops off as $\nu^{1/3}$ and at high frequencies ($\nu\gg\nu_m$) scales as
$\nu^{-(p-1)/2}$, where $p\approx 2.4$ is the power-law index for the 
energy distribution of electrons. 

If the energy in the magnetic field is taken to be some fraction, $\xi_B$, 
of the thermal energy density then $B^2/8\pi = \xi_B n_e\gamma_e m_e$. And 
thus the ratio of the peak synchrotron frequencies in regions III and II is

\begin{equation}
{\nu_{2m}\over \nu_{1m}} = \left({e_2\over e_1}\right)^{5/2} \left[{n_1
  \over n_2}\right]^2 = {(8\gamma_1^2-9)^{1.5}\over 3^{2.5} v_1^2},
\end{equation}
and the ratio of the observed flux from the outer shell after the forward
shock has traversed through it, and the flux in the absence of collision,
at a frequency much greater than the peak of the emission is given by

\begin{equation}
f_2(\nu) = \gamma_{f}' \left({e_2\over e_1}\right)^{1/2} 
   \left({\nu_{2m}\over \nu_{1m}}\right)^{(p-1)/2},
\end{equation}
where $\gamma_f'=\gamma_f(1+v_f v_0)$ is a factor by which the
Lorentz factor of the outer shell increases as a result of shell
collision. The dependence of the observed flux on $\gamma_f'$ can be
more rapid than the linear function considered above, depending on the
temporal profile of the shell acceleration.

At late times, when the collision has run its course, the shells have
merged and settled back to a BM solution, the observed flux is proportional
to $(E_1+E_2)^{(p+3)/4}$; in the absence of the shell collision
the flux would have been proportional to $E_1^{(p+3)/4}$.
Thus the increase in the  observed emission due to the forward shock
is approximately $(1+E_2/E_1)^{(p+3)/4}$. Numerical calculation of
flux increase, using equation (26), yields result that is consistent 
with this estimate.

The reverse shock traversing into the inner shell is mildly relativistic
but very strong. The density
enhancement in this case is about 4, i.e. $n_3/n_4 \sim 4$. Regions
IV and III are separated by a surface of contact discontinuity
across which the pressure is continuous but the density is not. The
thermal Lorentz factor of electrons in region IV, $\gamma_{t3}$,
can be calculated from the continuity of pressure at this interface 
and is given by

\begin{equation}
\gamma_{t3}= {e_3\over m_e n_3} = {3m_p\over 40 m_e}
  \left[{w_1 e_2\over w_4 e_1}\right],
\end{equation}
and the thermal Lorentz factor of electrons in region II, $\gamma_{t1}$,
can be shown to be equal to $(\gamma_{0c}/2^{3/2})(m_p/m_e)$ provided that 
electrons and protons are in equipartition;\footnote{The equipartition 
assumption is in fact not needed since we are only interested in
the ratio of peak frequencies and emissions from regions II, III, and IV.}
where $\gamma_{0c}\equiv \gamma_{0}(t_c)$ is the Lorentz factor of the
outer shell at the time of collision.
The ratio of the peak synchrotron frequencies in regions IV and II is
given by:

\begin{equation}
{\nu_{3m}\over \nu_{1m}} = {3^{1/2}\gamma_{t3}^2\over 2\gamma_{t1}^2}
\left({e_2 \over e_1}\right)^{1/2} \approx {1\over 7\gamma_{0c}^2} 
\left({e_2\over e_1}\right)^{5/2} \left({w_1\over w_4}\right)^2,
\label{stam}
\end{equation}
We have made use of the relation $B_3/B_1\approx (3e_2/4e_1)^{1/2}$ in 
deriving the above equation; $B_1$ \& $B_3$ are magnetic fields in 
regions II \& IV respectively. We have furthermore assumed that the 
fractional energy density in the magnetic field and electrons in 
regions IV and II are not different. This is an economical assumption 
which eliminates having to introduce additional free parameters.
Moreover, there is some observational support for this assumption
from the analysis of GRB970508 (Weijers \& Galma, 1998) and the
prompt optical flash of GRB990123 (sari \& Piran, 1999). It is
straightforward to modify the results presented here if this 
assumption is found to be invalid.

Using the result of \S2.1 we can write $e_2/e_1 = f(w_4/w_1)\approx
[1+(w_4/w_1)^{0.35}]$. Moreover, it can be easily shown that $w_4/w_1\approx
E_2/E_1$. Making use of all these results we find that the Lorentz factor 
of electrons in region IV is
smaller than the thermal Lorenz factor of electrons in the outer shell
by a factor of $\beta \sim 5 \gamma_{0c} \, (E_2/E_1)\,f^{-1}(E_2/E_1)$.
And so the peak frequency of the synchrotron emission emanating from the
shocked inner shell is smaller than the characteristic emission
frequency from the outer shell by a factor of $\beta^2 f^{-1/2}
\sim 7 \gamma_{0c}^2 (E_2/E_1)^{1.1}$.

The ratio of observed flux from the inner shell, after the reverse shock
has propagated through it, and the outer shell in the absence of
collision, at the peak frequency corresponding to the inner shell,
is given by:

\begin{equation}
{f_3(\nu_{3p})\over f_1(\nu_{3p})} = {\gamma_f' n_3 B_3\over n_1 B_1}
   \left[{\nu_{1m} \over\nu_{3m}}\right]^{1/3} = 
  {\gamma_f' n_3\over n_1} \left[ {B_3\gamma_{t1}\over B_1\gamma_{t3}}
  \right]^{2/3}
\end{equation}
Since $n_1=2^{3/2} \gamma_{0c} n_{ism}$, $w_1\approx 8\gamma_{0c}^2 n_{ism}
  m_p/3$, $n_3 \approx 4 n_4\approx 4 w_4/m_p$, we find that $n_3/n_1 \approx
\gamma_{0c} w_4/w_1$. Substituting this into the 
above equation we find that the observed emission from the inner 
shell at frequency $\gamma_{0c}\nu_3$,
is larger than the emission from the outer shell by a factor of
$\sim 8 [\gamma_{0c} E_2/E_1]^{5/3}$.

As an example consider that the inner shell is ejected with a Lorentz factor
of 5, and its energy is comparable to the energy in the outer shell
then the radiation from the shocked inner shell comes out at a frequency
which is smaller compared to the peak frequency of emission from the outer
shell by a factor of almost 10$^2$ and the observed flux at this frequency
is dominated by the inner shell by a factor of about 10$^2$; if the
energy in the inner shell were 1/5 of the energy in the outer shell then
the frequency ratio would be about 30 and the flux from the shocked 
inner shell larger by a factor of about 8.

Fig. 4 shows the peak frequency and the emission from the reverse shock,
and fig. 5 shows the modification to the light curve due to the forward 
shock.

The synchrotron cooling time in shell rest frame is $t_s = 8\pi c m_e/(\sigma_T
B^2 \gamma_e)$, and the dynamical time of the shell is $t_d = t/\gamma_0$.
Thus the ratio of the cooling to the dynamical time for the outer
shell is given by:

\begin{equation}
{t_s\over t_d} = {8\pi c m_e \gamma_0\over \sigma_T B^2 \gamma_e t} =
  {m_e^2\over \xi_e \xi_B \sigma_T c m_p^2 n_0 \gamma_0^2 t} \approx
  {1.58{\rm x}10^{-4} t_{obs}^{1/2}\over \xi_e\xi_B n_0^{1/2} E_{51}^{1/2}},
\end{equation}
where $\xi_e$ \& $\xi_B$ are the fraction of the total energy density
in electrons and the magnetic field. For $\xi_e\approx\xi_B=0.1$, $n_0=1$
and $E_{51} = 1$ the outer shell becomes adiabatic approximately one
hours after the burst in the observer's frame. The synchrotron cooling
time for the inner shell heated by the reverse shock is even longer
because the magnetic field strength in the regions III and IV are approximately
equal as a result of pressure balance across the surface of contact 
discontinuity (see fig. 1), and $\gamma_{3t}\ll\gamma_{2t}\sim 
\gamma_0$. Thus our estimate of energy flux calculated above under the 
assumption of adiabatic shock is valid as long as we restrict ourselves 
to shell collision after about one hour.

\section{Discussion and comparison with observations}

In the scenario that gamma-ray bursts are produced by internal shocks 
it is expected that some fraction of the energy of the explosion is 
carried by ejecta that is moving with moderate Lorentz factor which 
does not collide with faster moving material that was ejected at earlier
time until the faster shells have been slowed down by the ISM. We have 
explored the consequences of this possibility in this paper. 
The goal is to provide some constraint on the temporal behavior of the energy
release from the explosion.

When a slower moving shell hits a decelerating faster shell from behind,
it results in a correlated increase of emission at all frequencies. The 
relative Lorentz factor of collision can be shown to be about 1.25, and the 
amplitude of the enhanced emission depends on the ratio of the energy in 
the inner and the outer shells as well as the observed frequency.

The enhancement to the observed emission from the forward shock, that
propagates into the outer shell, is approximately $(1+E_2/E_1)^{1.4}$
and the characteristic synchrotron frequency is slightly larger than
the frequency of the outer shell in the absence of shell collision.

The emission from the reverse shock, which propagates into the inner
shell, is at a frequency that is smaller than the peak frequency of 
emission from the outer shell by a factor of $7\gamma_0^2(t_c)
(E_2/E_1)^{1.1}$; where $\gamma_0(t_c)$ is the Lorentz factor of the outer
shell at the time of collision. The emission
from the inner shell, at the peak frequency, is larger than the emission
from the outer shell at the same frequency by a factor of 
$\sim 8(\gamma_0 E_2/E_1)^{5/3}$.

Thus the observed spectrum, after shell collision, at a fixed time 
is expected to have two peaks. The lower frequency peak arises
from the inner shell and the higher frequency peak from the outer shell.
Moreover, the light curve at a fixed frequency has a bump that can be very
dramatic at low frequencies. The results presented in this paper are subject
to certain assumptions described in the introduction viz. the fraction
of the thermal energy carried by electrons and magnetic field is the
same in all shocks, and the thickness of the inner shell is of order
its radius divided by the square of its Lorentz factor which is expected
from causality considerations.

The observed light curve of the gamma-ray burst GRB970228 appears to have
a plateau during 6--10 days after the burst with an amplitude of
about 0.5 mag in the R-band (Fruchter et al. 1998). If this were to arise 
as a result of shell collision, one would conclude that the energy in the 
inner shell is about 40\% of the outer shell.

The light curve of the x-ray afterglow of GRB970508 shows a complicated
behavior. At about 4 days after the burst the light curve shows some deviation
from a power-law fall-off (Sokolov et al. 1997), from which
we infer the late time energy injection to be less than about 10\%
of the initial burst. At early times the behavior was more dramatic.
The observed x-ray flux in the 2--10 kev band fell between
11 and 16 hrs after the initial gamma-ray burst and then rose sharply
by almost a factor of two from about 16 to 20 hours. After a gap in the 
observation from 20 hours to 60 hours this was followed by a powerlaw decline
with time as $\sim t^{-1}$ (Piro et al. 1999; astrp-ph/9902013). A
possible interpretation of the sharp rise is that the outer decelerating
shell, which is producing the x-ray afterglow resulting from the 
shocked ISM, is being hit from behind by a shell with energy roughly
equal to the energy of the outer shell. If this is the correct interpretation
then we expect that the optical emission will follow this overall trend.
After an initial decline of flux that was seen in the early optical observation
(the earliest of which was about 5 hours after the burst) a sharp increase 
in the optical flux began one day after the GRB cf. Sokolov et al. (1997).
The optical flux peaked around two day after the burst and from then on
it began a powerlaw decline. It seems that there is a 
lag of almost a day between the optical and the X-ray light curves, which 
is not expected in the shell collision model as both the X-ray and 
the optical emissions are expected to arise from the forward shock (the 
emission from the reverse shock should peak in the milli-meter wavelength 
at this time, and at this wavelength the emission from the reverse shock
is larger than the emission from the forward shock by about two orders 
of magnitude). However the gaps in the  X-ray and optical observations 
do allow a much smaller lag or even no lag at all. If so than the jumps 
in the observed X-ray and optical fluxes  are consistant with enhanced 
emission due to a colliding shells (Panaitescu, Meszaros \& Rees, 1998) with 
a significant energy release. 

The detection of optical flash associated with GRB-990123 with peak
magnitude of 8.95 in the V-band 50 second after the initial gamma-ray
burst (Akerlof et al. 1999)  was remarkable. 
The recent work of Sari and Piran (1999) provides
a nice fit to the optical and radio observations based on this model.
It is interesting to note that these observations can also be
explained as a result of collision of two shells which collide with
a moderate relative Lorentz factor of order 2 or less. Such collisions
are expected in internal shock models when either a somewhat faster
moving shell takes over a slower shell or when a slower moving shell
catches up with a faster moving but decelerating shell. In the early,
radiative, phase of the shock the Lorentz factor of the outer shell
decreases with time as $1/t^3$, and in this case it can be easily shown 
that when slower moving shells catch up with a faster shell their 
relative Lorentz factor is 1.5. We have analyzed a simple model consisting 
of wind from the central source with randomly fluctuating Lorentz factor, 
lasting for about 100 sec (the duration of the GRB), and find that the 
total fluence in the 
optical wavelength band during the gamma-ray burst is about 0.1\% of 
the x-ray and the gamma-ray fluence, which is consistent with the 
observation reported by Akerlof et al. (1999).

The smoothness of the observed light curves of other GRB after-glow
emission suggests that the late time release of energy, in slower shells,
is typically quite small.

Future, multi-wave length, observations should provide more stringent
condition on the ejection of moderate/low Lorentz factor material from
the central engine of the gamma-ray bursts at late times.

This research was supported by the US-Israel BSF 95-328,
by a grant from the Israeli Space Agency and by a NASA grant
NAG5-3091. TP thanks Marc Kamionkowski and the Columbia Physics
Department for hospitality while this research was done, and Reem Sari 
and Shiho Kobayashi for helpful discussions.

\vfill\eject

\centerline{\bf REFERENCES}
\bigskip

\ni Akerlof, et al., 1999, Nature,  {\bf 398}, 400.

\ni Blandford, R.D. \& McKee, C.F. 1976, Phys. Fluids, 19, 1130

\ni Bond, H.E. 1997, IAU Circ. 6654

\ni Cohen, and Piran, T. 1999, ApJ., 518, 346. 

\ni Costa, E., et al. 1997, IAU Circ. 6572

\ni Frail, D.A. et al. 1997, Nature 389, 261

\ni Fruchter, A.S. et al. 1998, astro-ph/9801169

\ni Granot, J., Piran, T., \& Sari, R., 1998, ApJ  513, 679.

\ni Kabayashi, S., Piran, T. \& Sari, R. 1997, ApJ 490, 92

\ni Katz, J.I., \& Piran, T., 1997, ApJ 490, 772

\ni M\'esz\'aros, P., \& Rees, M.J. 1997, ApJ 476, 232

\ni M\'esz\'aros, P., Rees, M.J., \& Wijers, R. A. M. J., 1998, ApJ 499, 301

\ni Panaitescu, A., M\'esz\'aros, P. \& Rees, M.J. 1998, ApJ 503, 314

\ni Paczynski, B. 1997, Fourth Huntsville Gamma-Ray Burst Symposium

\ni Piran, T. 1999, Physics Reports, 314, 575. 

\ni Rees, M.J., \& M\'esz\'aros, P. 1998, ApJ 496, L1

\ni Sari, R., \& Piran, T. 1997a, MNRAS 287, 110

\ni Sari, R., \& Piran, T. 1997b, ApJ 485, 270

\ni Sari, R., \& Piran, T. 1999, ApJ 517, L109

\ni Sokolov et al. 1997, astro-ph/9709093

\ni van Paradijs, J., et. al. 1997, Nature 386, 686

\ni Vietri, M. 1997, ApJ 478, L9

\ni Wijers, R.A.M.J., \& Galma, T.J. 1998, astro-ph/9805341

\ni Wijers, R.A.M.J., Rees, M.J., \& M\'esz\'aros, P. 1997, MNRAS, 288, L51

\vfill\eject
\centerline{\bf Figure Captions}
\bigskip

\ni Figure 1.--- The panel to the left shows the Lorentz factor of ejected
shells as a function of time (random function). The panel to the right
shows the Lorentz-factor-distribution as a function of radial distance 
from the center at some time when the slower moving shells have been 
taken over by faster shells leading to a monotonically increasing 
Lorentz factor with distance (a constant was subtracted from the distance).
The width of lines is proportional to shell mass. 

\ni Figure 2.--- Shows the various shock regions when a cold
shell collides with a hot shell.

\medskip
\ni Figure 3.--- The graph show the ratio of the thermal energy density
in regions (III) and II ($e_2/e_1$) as a function of the ratio of the 
enthalpy density of the cold inner shell, region V, and the outer shell,
region II, ($w_4/w_1$); the ratio $w_4/w_1$ is approximately equal to 
the ratio of energy in the inner and the outer shells. The four curves 
corresponds to four different values of the relative Lorentz factor
of the two shells ($\gamma_{14}$): the value of $\gamma_{14}$ for the
bottom to the top curves are 1.25, 1.5, 2.0 and 3.0 respectively.
Note that the minimum value of $w_4/w_1$ for which there is a forward
shock i.e. $e_2/e_1>1$, decreases with increasing $\gamma_{14}$. The
value of $\gamma_{14}$ is close to 1.25 for the case where a slow moving
shell catches with a faster but decelerating shell.

\medskip
\ni Figure 4.--- The top panel shows the ratio of the peak synchrotron 
frequency in the region of the reverse shock to the outer shell multiplied
by $\gamma_{0c}^2$ ($\gamma_{0c}$ is the thermal Lorentz factor of gas
in the outer shell or region II). The lower panel shows the ratio of 
emission from the reverse shock region and the outer shell, at the 
characteristic synchrotron frequency of region IV, divided by $\gamma_{0c}^{
5/3}$, as a function of $w_4/w_1\approx E_2/E_1$ at a time when the 
reverse shock has reached the back end of the inner shell.
The solid curve is for $\gamma_{14} = 1.25$ and the
dotted curve is for $\gamma_{14}=2.0$. The horizontal part of the curve
for $\gamma_{14} = 1.25$ corresponds to the forward shock stalling at
a point in the outer shell where $w_4/w_1=0.36$, and the physical parameters
in the reverse shock is determined by this critical point.

\medskip
\ni Figure 5.--- The effect of shell collision on the afterglow
light curve (schematic). The continuous curve corresponds to $E_2/E_1=0.4$
and the dashed curve is for $E_2/E_1=0.7$

\end{document}